\documentclass[11pt]{article}
\begin{document}

\title{{\bf Supersymmetric Moyal-Lax Representation}}
\author{Ashok Das\\
Department of Physics and Astronomy,\\
University of Rochester,\\
Rochester, NY 14627-0171\\
USA\\
\\
and\\
\\
Ziemowit Popowicz \\
Institute of Theoretical Physics, \\
University of Wroc\l aw,\\
50-205 Wroc\l aw\\ 
Poland.}
\date{}
\maketitle

\begin{abstract}

The super Moyal-Lax representation and the super Moyal momentum
algebra are  introduced and the properties of simple and extended
supersymmetric integrable models are  systematically investigated.
It is shown that, much like in the bosonic cases, the super Moyal-Lax
equation can be interpreted as a Hamiltonian equation and can be derived  
from an action. Similarly, we show that the parameter of non-commutativity, in
this case, is related to  the central charge 
of the second Hamiltonian structure of the system. The super Moyal-Lax
description allows us to go to the dispersionless limit of these
models in a singular limit and we discuss some of the properties of
such systems.

\end{abstract}

\newpage

\section{Introduction:}

Integrable models \cite{1}, both bosonic as well as supersymmetric
\cite{2}-\cite{6},  have played
important roles in the study of conformal field theories, strings,
membranes and topological field theories. In recent years, it has
become known that string (membrane) theories naturally lead to
non-commutative field theories \cite{7}, where usual multiplication of
functions is replaced by the star product of Groenewold \cite{8} and
Moyal \cite{9}. It is known now that Moyal brackets can be used in soliton
theory as well \cite{10}-\cite{12}. In an earlier paper \cite{13}, we
constructed the Moyal-Lax representation for bosonic integrable
models, using the star product of Groenewold. Such an approach has
some very attractive features. For example, a
Moyal-Lax equation can be given the meaning of a Hamiltonian equation
and can be derived from an action. We also showed that a Moyal-Lax
equation naturally leads to the appropriate Lax equation, in terms of
Poisson brackets, in the dispersionless limit. The Moyal-Lax equation
was also shown to lead in a simple manner to the Hamiltonian structures
(at least the first two) of the dispersionless systems, which 
was an open problem for quite some time.

In this paper, we continue our investigation of the star product and
the Moyal-Lax representation for supersymmetric integrable
models. These are models defined on a super space \cite{14} and we adopt the
supersymmetric star product \cite{15} as well as the Moyal bracket to
construct and show that a consistent Moyal-Lax representation for
supersymmetric
integrable systems can, in fact, be obtained. Much like the bosonic
case, the supersymmetric Moyal-Lax representation can be given the
meaning of a Hamiltonian equation and can be derived from an action in
superspace. The dispersionless limit of supersymetric systems is
problematic in general. We show that it is possible to obtain the
dispersionless models from such a representation in a singular limit,
which, however, is not very practical from a calculational point of
view. Therefore, the alternate Lax descriptions obtained in the
literature \cite{16}-\cite{17}are still preferable. However, as yet,
there  is no
systematic understanding of how to obtain such alternate Lax
representations. The paper is organized as follows. In section {\bf
2}, we describe the basic star product and the Moyal bracket
generalized to superspace. We also present various other identities
that are useful in deriving the Moyal-Lax representation for
supersymmetric integrable systems. In section {\bf 3}, we describe, in
detail, the standard Moyal-Lax representation for $N=1$ supersymmetric
KdV hierarchy. We show how this equation can be thought of as a
Hamiltonian equation, which can be derived from an action. We also
point out how the dispersionless limit of this system can be obtained
in a singular limit. In section {\bf 4}, we discuss several other
models with $N=1$ supersymmetry, both in standard and the non-standard
representations, as examples
(without going into too much details). In particular, we describe the
non-standard representation for the supersymmetric KdV equation (sKdV),
supersymmetric two boson system (sTB), supersymmetric non-linear
Schr\"{o}dinger equation (sNLS) and the supersymmetric modified KdV (smKdV)
equation.  In section {\bf 5}, we consider the
Moyal-Lax representation for systems with extended supersymmetry. We
discuss the $N=2$ sKdV systems and bring out various properties of
these  systems from  the Moyal-Lax representation. In section {\bf 6},
we present a brief conclusion. All the calculations presented in this
paper have also been checked using REDUCE \cite{21} and package Susy2
\cite {22}.

\section{Basic relations:}

In a bosonic phase space, the star product of two observables is
defined to be
\begin{equation}
A(x_{i},p_{i})\star B(x_{i},p_{i}) = e^{\kappa\sum_{i=1}^{n}
(\partial_{x_{i}}\partial_{\tilde{p}_{i}}-\partial_{p_{i}}
\partial_{\tilde{x}_{i}})} 
\left.A(x_{i},p_{i})B(\tilde{x}_{i},\tilde{p}_{i})\right|_{\tilde{x}_{i}=x_{i},
\tilde{p}_{i}=p_{i}}
\end{equation}
where $n$ represents the number of coordinates. Here, in principle,
the  deformation parameter $\kappa$ can be
different along different directions. However, if we impose rotational
invariance, they can all be identified. In dealing with
supersymmetric systems, on the other hand, the
natural phase space manifold is a graded manifold with coordinates
$x_{i},p_{i},\theta_{\alpha},p_{\theta_{\alpha}}$, where
$\theta_{\alpha}$ are the fermionic coordinate
and $p_{\theta_{\alpha}}$ the corresponding conjugate momenta. Here,
we have taken a very general set up because we will be discussing
systems with simple supersymmetry as well as ones with extended
supersymmetry.  The Grassmann
variables satisfy anti-commutation relations. Consequently, they are
nilpotent and the derivatives with respect to such variables are
directional. In our discussions, we will use a left derivative for the
Grassmann variables. Denoting the phase space variables collectively
as $z_{A} = (x_{i},p_{i},\theta_{\alpha},p_{\theta_{\alpha}})$, we
note that the star product can be
generalized to such a phase space, which is a graded manifold, as
\cite{15} 
\begin{equation}
A(z_{A})\star B(z_{A}) =
e^{\kappa\left[\sum_{i=1}^{n}(\partial_{x_{i}}\partial_{\tilde{p}_{i}}-
\partial_{p_{i}}\partial_{\tilde{x}_{i}}) 
+ \sum_{\alpha =1}^{N}
(\partial_{\theta_{\alpha}}\partial_{\tilde{p}_{\theta_{\alpha}}}+
\partial_{p_{\theta_{\alpha}}}
\partial_{\tilde{\theta}_{\alpha}})\right]}
\left. A(z_{A})B(\tilde{z}_{A})\right|_{\tilde{z}_{A}=z_{A}}\label{1}
\end{equation}
where $n$ represents the number of bosonic coordinates while $N$
corresponds to the number of fermionic (Grassmann) coordinates of the
manifold. 
The relative sign of the fermionic derivative terms, as we will see,
is chosen so as to bring out the Poisson brackets, on such a graded
manifold with a left derivative, correctly \cite{18}-\cite{19}. (After
all,  the star
product can be thought of as the exponentiation of the Poisson bracket
structure.)

Let us consider, in detail, the properties of such a star product in
the case of a simple graded manifold with one bosonic and one
fermionic coordinate, namely, $n=1=N$. With the star product defined as in
Eq. (\ref{1}), it is easy to verify that
\begin{eqnarray}
x\star x & = & x^{2},\quad p\star p = p^{2},\quad \theta\star \theta =
\theta^{2} = 0,\quad p_{\theta}\star p_{\theta} = p_{\theta}^{2} =
0\nonumber\\
x\star \theta & = & x\theta = \theta\star x,\quad x\star p_{\theta} =
xp_{\theta} = p_{\theta}\star x,\quad \theta\star p = \theta p =
p\star \theta\nonumber\\
x\star p & = & xp + \kappa,\quad p\star x = px - \kappa,\quad \theta\star
p_{\theta} = \theta p_{\theta} - \kappa,\quad p_{\theta}\star \theta =
p_{\theta}\theta - \kappa
\end{eqnarray}
Since, on a superspace, we can have both even and odd functions
(superfields), the graded Moyal bracket of these superfields can be
defined to be
\begin{equation}
\left\{A,B\right\}_{\kappa} = {1\over 2\kappa} \left(A\star B -
(-1)^{|A||B|} B\star A\right)\label{2}
\end{equation}
where $|A|,|B|$ represent the Grassmann parity of the superfields
$A,B$ respectively. It can be easily checked that with Eq. (\ref{2}),
we obtain
\begin{equation}
\left\{x,p\right\}_{\kappa} = 1 = -
\left\{\theta,p_{\theta}\right\}_{\kappa}\label{3}
\end{equation}
with all other graded Moyal brackets vanishing. 

It is also easy to check from the definition in Eq. (\ref{2}) that, in
the limit of vanishing $\kappa$, they lead to the correct definitions
of Poisson brackets on a graded manifold. Namely, if we assume that
$B, F$ represent respectively a bosonic and a fermionic superfield,
then, it follows from Eqs. (\ref{1},\ref{2}) that
\begin{eqnarray}
\lim_{\kappa\rightarrow 0} \left\{B_{1},B_{2}\right\}_{\kappa} & = &
\partial_{x}B_{1}\partial_{p}B_{2}-\partial_{p}B_{1}\partial_{x}B_{2}+
\partial_{\theta}B_{1}\partial_{p_{\theta}}B_{2}+
\partial_{p_{\theta}}B_{1}\partial_{\theta}B_{2} =
\left\{B_{1},B_{2}\right\}\nonumber\\
\lim_{\kappa\rightarrow 0} \left\{B_{1},F_{2}\right\}_{\kappa} & = &
\partial_{x}B_{1}\partial_{p}F_{2}-\partial_{p}B_{1}\partial_{x}F_{2}+
\partial_{\theta}B_{1}\partial_{p_{\theta}}F_{2}+
\partial_{p_{\theta}}B_{1}\partial_{\theta}F_{2} =
\left\{B_{1},F_{2}\right\}\nonumber\\
\lim_{\kappa\rightarrow 0} \left\{F_{1},B_{2}\right\}_{\kappa} & = &
\partial_{x}F_{1}\partial_{p}B_{2}-\partial_{p}F_{1}\partial_{x}B_{2}-
\partial_{\theta}F_{1}\partial_{p_{\theta}}B_{2}-
\partial_{p_{\theta}}F_{1}\partial_{\theta}B_{2} =
\left\{F_{1},B_{2}\right\}\nonumber\\
\lim_{\kappa\rightarrow 0} \left\{F_{1},F_{2}\right\}_{\kappa} & = &
\partial_{x}F_{1}\partial_{p}F_{2}-\partial_{p}F_{1}\partial_{x}F_{2}-
\partial_{\theta}F_{1}\partial_{p_{\theta}}F_{2}-
\partial_{p_{\theta}}F_{1}\partial_{\theta}F_{2} =
\left\{F_{1},F_{2}\right\}\label{4}
\end{eqnarray}
These are, in fact, the correct definitions of Poisson brackets on a
graded manifold with a left derivative \cite{18}-\cite{19}. It follows
from  this, as well
as from Eq. (\ref{3}) that in the limit $\kappa\rightarrow 0$, we have
the expected canonical Poisson bracket relations
\begin{equation}
\left\{x,p\right\} = 1 = - \left\{\theta,p_{\theta}\right\}
\end{equation}
with all others vanishing.

On a simple superspace of the kind we are considering, one can define,
in addition to the usual bosonic and fermionic derivatives, a covariant
derivative which transforms covariantly under supersymmetry
transformations, namely,
\begin{equation}
D = \partial_{\theta} + \theta \partial_{x}\label{5}
\end{equation}
Furthermore, the covariant derivative satisfies the relation that
\begin{equation}
D^{2} = \partial_{x}
\end{equation} 
We note here that we can define a fermionic quantity, from the phase
space variables, as
\begin{equation}
\Pi = -\left(p_{\theta}+\theta\star p\right) = - \left(p_{\theta} +
\theta p\right)\label{6}
\end{equation}
which would satisfy
\begin{equation}
\Pi\cdot\Pi = 0,\qquad \Pi\star \Pi = - 2\kappa p\label{7}
\end{equation}
It is easy to check now that, independent of the Grassmann parity of a
superfield $A$, we have
\begin{equation}
\left\{\Pi,A\right\}_{\kappa} =  (DA)
\end{equation}
Namely, the graded Moyal bracket of $\Pi$ with any superfield leads to
the covariant derivative acting on the superfield for any value of
$\kappa$  (even in the
vanishing $\kappa$ limit). This is, therefore, an important concept in
the study of supersymmetric integrable systems. In fact, one can think
of this as the generalization of the fermionic momentum variable
$p_{\theta}$ to one which is covariant with respect to supersymmetric
transformations (translations of the Grassmann coordinates).

From the definition of the star product, it is now easy to check that,
for any integer $n$ (positive or negative)
\begin{equation}
p^{n}\star A = \sum_{m=0} \left(\begin{array}{c} n\\
m\end{array}\right) (-2\kappa)^{m}\left({\partial^{m}A\over \partial
x^{m}}\right) \star p^{n-m}\label{8}
\end{equation}
where
\[
\left(\begin{array}{c} n\\ m\end{array}\right) = {n(n-1)\cdots
(n-m+1)\over m!},\qquad \left(\begin{array}{c} n\\ 0\end{array}\right)
= 1
\]
Similarly, it can be checked that (powers of $\Pi$ are defined in the star
product sense)
\begin{eqnarray}
\Pi^{2n}\star A & = & \sum_{m=0} \left(\begin{array}{c} n\\
m\end{array}\right) (-2\kappa)^{2m} (D^{2m}A)\star
\Pi^{2(n-m)}\label{9}\\
\Pi^{2n+1}\star A & = & \sum_{m=0} \left(\begin{array}{c} n\\
m\end{array}\right) (-2\kappa)^{2m}\left((-1)^{|A|}(D^{2m}A)\star
\Pi^{2(n-m)+1} + (2\kappa) (D^{2m+1}A)\star
\Pi^{2(n-m)}\right)\nonumber
\end{eqnarray}
These can be thought of as the generalizations of the super-Leibniz
rules \cite{5} to the case of the star product on a superspace.

For completeness, let us note that
\[
\Pi\star p = \Pi p = p\star \Pi 
\]
and that, for a non-vanishing $\kappa$, we can define, from Eq. (\ref{7}),
\begin{equation}
\Pi^{-1} = (-2\kappa)^{-1} \Pi\star p^{-1} = (-2\kappa)^{-1}\Pi p^{-1}
= (-2\kappa)^{-1} p^{-1}\star \Pi\label{10}
\end{equation}
This inverse, on the other hand, does not exist in the vanishing $\kappa$
limit. As we will see later, this is one of the sources of
difficulties in taking the dispersionless limit of supersymmetric
integrable systems.

\section{Supersymmetric KdV hierarchy:}

On the phase space of a supersymmetric system, which is a graded
manifold, with a star product, we can define a Lax function which 
depends on the phase space coordinates as well as on dynamical
variables which will be superfields (either bosonic or fermionic). In
fact, for a manifestly supersymmetric description, the Lax function
can depend only on $p,\Pi$ as well as on superfields and 
covariant derivatives acting on them. Thus, for example, a  
Lax operator can have a form of the type
\begin{equation}
L_{n} = \sum_{m=0} \Phi_{m}(x,\theta)\star p^{n-m}\Pi^{m}\label{11}
\end{equation}
where all products are star products (for example,
$\Pi^{m}=\Pi\star\cdots \star \Pi$ with $m$ factors). Although such
Lax  operators are defined as polynomials in momenta,
they inherit an operator structure through the star product and define
an algebra, which we will call the super-Moyal momentum algebra (sMm
algebra). It is easy to check that all the properties of pseudo
differential operators on a superspace carry through, in this case,
with suitable redefinitions. In particular, we note that for any two
arbitrary elements $A$ and $B$ of the sMm algebra, the super residue
(the coefficient of the $\Pi^{-1}$ term) of the super Moyal bracket
satisfies
\begin{equation}
sRes\,\left\{A,B\right\}_{\kappa} = (DC)
\end{equation}
so that we can define uniquely a super trace
\begin{equation}
sTr\,A = \int dx\,d\theta\;sRes\,A\label{11'}
\end{equation}
which will satisfy cyclicity.

For a general Lax operator of the type in Eq. (\ref{11}), one can
readily check that a super Moyal-Lax equation of the form
\begin{equation}
{\partial L_{n}\over \partial t_{k}} = \left\{L_{n}, \left(L_{n}^{k\over
n}\right)_{\geq m}\right\}_{\kappa},\qquad k\neq ln
\end{equation}
where $k,l$ are integers and $()_{\geq m}$ denotes the projection with
respect to powers of $\Pi$ with the star product, defines a consistent
Lax equation only if $m=0,1,2$. Note here that for any element $A$ of the
sMm algebra, $A^{k\over n} = A^{1\over n}\star A^{1\over n}\star\cdots
\star A^{1\over n}$, where the $n$th root is determined formally in a
recursive manner. The projection with $m=0$ is conventionally denoted
as $()_{+}$ and an equation with the projection $m=0$ is called a
standard super Moyal-Lax representation while for the other
projections, the equations are known as non-standard representations.

Let us describe in detail how all of this works in the case of the
$N=1$ supersymmetric KdV hierarchy. Let us consider a fermionic
superfield of the form
\begin{equation}
\Phi(x,\theta) = \psi(x) + \theta u(x)
\end{equation}
and a Lax function which is an element of the sMm algebra of the form
\begin{equation}
L = p^{2} + \Pi\star \Phi = p^{2} - \Phi\star \Pi + 2\kappa (D\Phi)
\end{equation}
In this case, we can show in a straightforward manner (all
products and projections are with respect to star product) that
\begin{equation}
\left(L^{3\over 2}\right)_{+} = p^{3} - {3\over 2}\Phi\star p\Pi +
3\kappa (D\Phi)\star p + {3\kappa\over 2} \Phi_{x}\star \Pi -
3\kappa^{2} (D\Phi_{x})
\end{equation}
where the subscript $x$ stands for a derivative with respect to the
space coordinate. It is tedious, but straightforward to check that the
super Moyal-Lax equation (in the standard representation)
\begin{equation}
{\partial L\over \partial t} = {2\over \kappa}\,\left\{L, \left(L^{3\over
2}\right)_{+}\right\}_{\kappa}\label{12}
\end{equation}
leads to
\begin{equation}
{\partial\Phi\over \partial t} = \left(2\kappa \Phi_{xx} + 3 \Phi
(D\Phi)\right)_{x}\label{13}
\end{equation}
which we recognize as the $N=1$ supersymmetric KdV equation \cite{2}
(The conventional representations of the equation corresponds to
$2\kappa=1$).  In fact,
the entire supersymmetric KdV hierarchy can be obtained from a super
Moyal-Lax equation of the form (up to a multiplicative constant)
\begin{equation}
{\partial L\over \partial t_{k}} = \left\{L, \left(L^{2k+1\over
2}\right)_{+}\right\}_{\kappa}\label{14}
\end{equation}

Let us next show that the super Moyal-Lax equation of
(\ref{14}) can be derived from an action and can be given the meaning
of a Hamiltonian equation. Let us consider a phase space action of
the form
\begin{equation}
S = \int dt \left(p\star \dot{x} + \dot{\theta}\star p_{\theta} -
\left(L^{2k+1\over 2}\right)_{+}\right)
\end{equation}
Here the particular ordering of the velocity in the second term
reflects our choice of a left derivative \cite{18}-\cite{19}. It can
now be  easily checked
that the Euler-Lagrange equations following from this action lead to
\begin{eqnarray}
\dot{x} & = & {\partial \left(L^{2k+1\over 2}\right)_{+}\over \partial
p} = \left\{x,\left(L^{2k+1\over
2}\right)_{+}\right\}_{\kappa}\nonumber\\
\dot{p} & = & -  {\partial \left(L^{2k+1\over 2}\right)_{+}\over \partial
x} = \left\{p,\left(L^{2k+1\over
2}\right)_{+}\right\}_{\kappa}\nonumber\\
\dot{\theta} & = & - {\partial \left(L^{2k+1\over 2}\right)_{+}\over \partial
p_{\theta}} = \left\{\theta ,\left(L^{2k+1\over
2}\right)_{+}\right\}_{\kappa}\nonumber\\
\dot{p}_{\theta} & = & - {\partial \left(L^{2k+1\over
2}\right)_{+}\over \partial \theta} = \left\{p_{\theta},\left(L^{2k+1\over
2}\right)_{+}\right\}_{\kappa}
\end{eqnarray}
Namely, these are the appropriate Hamiltonian equations for the system
with the super Moyal bracket playing the role of the Poisson bracket
and $\left(L^{2k+1\over 2}\right)_{+}$ representing the
Hamiltonian. The dynamical evolution of any other variable can now be
obtained in a simple manner and, in particular, we note that
\begin{equation}
{\partial L\over \partial t_{k}} = \left\{L, \left(L^{2k+1\over
2}\right)_{+}\right\}_{\kappa}
\end{equation}
This shows that the super Moyal-Lax equation can indeed be thought of
as a Hamiltonian equation with $\left(L^{2k+1\over 2}\right)_{+}$ playing
the role of the Hamiltonian, much like in bosonic integrable systems
\cite{13}.  Although we have shown this explicitly
for a super Moyal-Lax equation in the standard representation, it is
quite clear that this derivation generalizes to a super Moyal-Lax
equation  with a non-standard representation as well as systems
with extended supersymmetry.

The conserved quantities of the system can be obtained in a simple
manner. Using the definition of {\it super trace} in Eq. (\ref{11'}),
we can write (up to a multiplicative constant)
\begin{equation}
H_{2m+1} = -{1\over (2m+1) \kappa^{m}}\;sTr\,L^{2m+1\over 2}\label{15'}
\end{equation}
and expressing the Hamiltonians in terms of densities as
\[
H_{2m+1} = \int dx\,d\theta\,{\cal H}_{2m+1}
\]
the first few Hamiltonian densities take the forms
\begin{eqnarray}
{\cal H}_{1} & = & \Phi\nonumber\\
{\cal H}_{3} & = & {1\over 4}\;\Phi (D\Phi)\nonumber\\
{\cal H}_{5} & = & {1\over 4}\left(\kappa \Phi(D\Phi_{xx}) + \Phi
(D\Phi)^{2}\right)\nonumber\\
{\cal H}_{7} & = & {1\over 16}\left[4\kappa^{2} \Phi(D\Phi_{xxxx}) +
2\kappa \left(\Phi_{xx}\Phi_{x}\Phi + 7 \Phi(D\Phi_{xx})(D\Phi) + 4
(D\Phi)^{2}\right) + 5 \Phi(D\Phi)^{3}\right]\label{15}
\end{eqnarray}
and so on. It is easy to check that these quantities are conserved
under the flow of the sKdV hierarchy.

The discussion of the Hamiltonian structures for the sKdV hierarchy can
be carried out much along the lines of pseudo differential
operators. Since it is rather technical, we simply give the results
here. The first Hamiltonian structure is highly non-local and has the
form
\begin{equation}
{\cal D}_{1} = 8\kappa D^{2} (2\kappa D^{3}+\Phi)^{-1} D^{2}\label{16}
\end{equation}
while the second structure, corresponding to the superconformal
algebra, has the form
\begin{equation}
{\cal D}_{2} = 2 \left(2\kappa D^{5} + 3\Phi D^{2} + (D\Phi) D + 2
(D^{2}\Phi)\right)\label{17}
\end{equation}
so that the sKdV equation, Eq. (\ref{13}), can be written as
\begin{equation}
{\partial \Phi\over \partial t} = {\cal D}_{1} {\delta H_{5}\over
\delta \Phi} = {\cal D}_{2} {\delta H_{3}\over \delta \Phi}
\end{equation}
From the structure in (\ref{17}), it is also clear that the
non-commutativity parameter, $\kappa$, is related to the central
charge of the superconformal algebra, much as we had shown earlier
\cite{13} 
that in a bosonic integrable model, it is related to the central
charge in the algebra of the second Hamiltonian structure.

Let us note here that in the limit $\kappa\rightarrow 0$,
Eq. (\ref{13}) does lead to the dispersionless limit of the $N=1$
supersymmetric KdV equation \cite{16}, namely,
\begin{equation}
{\partial \Phi\over \partial t} = 3 \left(\Phi(D\Phi)\right)_{x}
\end{equation}
However, we note from Eq. (\ref{12}) that this is obtained in a
singular limit which is not very amenable to manipulations. It is for
this reason that an alternate Lax representation for the
dispersionless equation has proven much more useful \cite{16} and the origin of
such an alternate Lax representation remains an open question. It is
also worth noting here that, in the limit $\kappa\rightarrow 0$, the
conserved quantities in Eq. (\ref{15'},\ref{15}) do reduce to the
local conserved quantities of the dispersionless sKdV
hierarchy. Furthermore, in this limit, the second Hamiltonian
structure in Eq. (\ref{17}) reduces to the center less superconformal
algebra, which is known to be a Hamiltonian structure of the
dispersionless model. On the other hand, the first Hamiltonian
structure in Eq. (\ref{16}) vanishes in this limit, which explains why
such a structure has not been found within the context of the
dispersionless model. Without going into details, we will like to note
here that the super Moyal-Lax representation also allows us to
construct the non-local charges of the $N=1$ supersymmetric KdV
system, which reduces in the dispersionless limit to one of the two
sets of non-local charges found in the literature \cite{16}. The
understanding of the non-local charges, within the context of
dispersionless supersymmetric  systems, therefore, remains an open
question. 

\section{Other examples:}

In this section, we will discuss briefly the super Moyal-Lax
representations for some other integrable models with $N=1$
supersymmetry. 

a) {\it Nonstandard KdV}:
\medskip

$N=1$ supersymmetric KdV can also be given a non-standard description
as follows. Let us consider the Lax operator
\begin{equation}
L = p + p^{-1}\star \Pi\star \Phi
\end{equation}
Then, it is easy to check that (projection with respect to the star
product and powers of $\Pi$),
\begin{equation}
\left(L^{3}\right)_{\geq 1} = p^{3} + 6\kappa (D\Phi)\star p -
3\Phi\star p\star \Pi
\end{equation}
It follows from this that the super Moyal-Lax equation
\begin{equation}
{\partial L\over \partial t} = {1\over 2\kappa} \left\{L,
\left(L^{3}\right)_{\geq 1}\right\}_{\kappa}
\end{equation}
leads to the $N=1$ susy KdV equation
\begin{equation}
{\partial \Phi\over \partial t} = \left(4\kappa^{2} \Phi_{xx} + 3 \Phi
(D\Phi)\right)_{x}
\end{equation}
Once again, we see that the dispersionless limit can be obtained in
the singular limit, $\kappa\rightarrow 0$.

b) {\it Supersymmetric two boson equation}:
\medskip

The supersymmetric two boson equation \cite{5} also has a non-standard super
Moyal-Lax representation of the following form. Let us consider the
Lax operator of the form
\begin{equation}
L = p - (D\Phi_{0}) + \Pi^{-1}\star \Phi_{1}
\end{equation}
Here, both $\Phi_{0}$ and $\Phi_{1}$ are considered to be fermionic
super fields. Then, it follows that
\begin{equation}
\left(L^{2}\right)_{\geq 1} = p^{2} - 2 (D\Phi_{0})\star p +
(2\kappa)^{-1} \Phi_{1}\star \Pi
\end{equation}
and the super Moyal-Lax equation
\begin{equation}
{\partial L\over \partial t} = \left\{L, \left(L^{2}\right)_{\geq
1}\right\}_{\kappa}
\end{equation}
leads to the consistent Hamiltonian equations
\begin{eqnarray}
{\partial \Phi_{0}\over \partial t} & = & -\left(2\kappa \Phi_{0xx} +
(D(D\Phi_{0})^{2}) + 2\Phi_{1x}\right)\nonumber\\
{\partial \Phi_{1}\over \partial t} & = & \left(2\kappa \Phi_{1x} + 2
\Phi_{1} (D\Phi_{0})\right)_{x}
\end{eqnarray}
These equations are easily seen to reduce to the correct
dispersionless system \cite{17} in the limit $\kappa\rightarrow
0$. Let us  also
note here that the supersymmetric KdV hierarchy is embedded in the
supersymmetric two boson hierarchy (with $\Phi_{0}=0$).

c) {\it Supersymmetric non-linear Schr\"{o}dinger equation}:
\medskip

It is known that the supersymmetric two boson equation is related to
the supersymmetric non-linear Schr\"{o}dinger equation through a field
redefinition \cite{5}. Let us next show that this holds even with a Moyal-Lax
representation, which clarifies some of the features of the
dispersionless limit of this model. Let us define
\begin{equation}
\Phi_{0} = 2\kappa (D \ln (DQ)) + (D^{-1}(\overline{Q}Q)),\quad
\Phi_{1} = 2\kappa \overline{Q} (DQ)
\end{equation}
With this redefinition, the Lax operator for the susy two boson
hierarchy becomes
\begin{eqnarray}
L & = & p - (D\Phi_{0}) + \Pi^{-1}\star \Phi_{1}\nonumber\\
 & = & p - 2\kappa {(D^{3}Q)\over (DQ)} - \overline{Q}Q + 2\kappa
 \Pi^{-1}\star \overline{Q} (DQ)\nonumber\\
 & = & (DQ)^{-1}\star \left(p - \overline{Q}Q + 2\kappa (DQ)^{-1}\star
 \Pi^{-1}\star \overline{Q}\right)\star (DQ)\nonumber\\
 & = & G\star \tilde{L}\star G^{-1}
\end{eqnarray}
One says that the Lax operators $L$ and $\tilde{L}$ are related through
a gauge transformation $G$.

It is easy to verify that the new Lax operator $\tilde{L}$ does not
lead to any consistent Moyal-Lax equation. However, let us define (in
the language of pseudo differential operators, this will be called a
formal adjoint)
\begin{equation}
{\cal L} = \tilde{L}^{T} = - \left(p + \overline{Q}Q - 2\kappa
\overline{Q}\star \Pi^{-1}\star (DQ)\right)\label{a}
\end{equation}
Then, the super Moyal-Lax equation
\begin{equation}
{\partial {\cal L}\over \partial t} = {1\over 4\kappa} \left\{{\cal
L}, \left({\cal L}^{2}\right)_{\geq 1}\right\}_{\kappa}
\end{equation}
leads to
\begin{eqnarray}
{\partial Q\over \partial t} & = &  \kappa Q_{xx} + 
\left(D((DQ)\overline{Q}))\right) Q\nonumber\\
{\partial \overline{Q}\over \partial t} & = & - \kappa\overline{Q}_{xx} -
\left(D((D\overline{Q}) Q)\right) \overline{Q}
\end{eqnarray}
which are the supersymmetric non-linear Schr\"{o}dinger
equations. This shows that a dispersionless limit of this set of
equations can be obtained in the singular $\kappa\rightarrow 0$ limit
(note the powers of $\kappa$ in the field redefinitions as well),
which explains why a direct construction of such a model has not
succeeded so far.

d) {\it Supersymmetric mKdV equation}:

We also note here that, if we make the identification $\overline{Q}=Q$
(recall that $Q$ is fermionic), then the Lax operator in Eq. (\ref{a})
will define a consistent super Moyal-Lax equation
\begin{equation}
{\partial {\cal L}\over \partial t} = {1\over 2\kappa} \left\{{\cal
L}, \left({\cal L}^{3}\right)_{\geq 1}\right\}_{\kappa}
\end{equation}
leading to
\begin{equation}
{\partial Q\over \partial t} = 2\kappa Q_{xxx} - 3 (D^{2}(Q(DQ)) (DQ)
\end{equation}
We recognize this to be the $N=1$ supersymmetric mKdV (modified KdV)
equation. In the limit $\kappa\rightarrow 0$, this leads to the
correct dispersionless limit (be it in a singular manner)
\begin{equation}
{\partial Q\over \partial t} = - 3 (D^{2}(Q(DQ)) (DQ)
\end{equation}
whose bosonic limit yields the dispersionless limit of the mKdV
equation (which is a higher order flow of the Riemann hierarchy).  

\section{Examples of systems with extended supersymmetry:}
\medskip

In this section, we will discuss the distinct $N=2$
supersymmetrizations of the KdV equation as examples of systems with
extended supersymmetry. Let us note that the natural setting for a
description of the $N=2$ supersymmetric KdV
hierarchies is the $N=2$ superspace, which is parameterized by two
fermionic (Grassmann) coordinates $\theta_{1},\theta_{2}$ in addition
to the usual bosonic coordinate $x$ (In the notation of section {\bf
2}, $n=1,N=2$). In this case, there are two
possible covariant derivatives that one can define, namely,
\begin{equation}
D_{1} = \partial_{\theta_{1}} + \theta_{1} \partial_{x},\qquad D_{2} =
\partial_{\theta_{2}} + \theta_{2} \partial_{x}
\end{equation}
These covariant derivatives satisfy the algebraic relations
\begin{equation}
D_{1}^{2} = D_{2}^{2} = \partial_{x},\qquad D_{1}D_{2} = - D_{2}D_{1}
\end{equation}
Correspondingly, on the phase space manifold of such a system, we can
define two variables
\begin{equation}
\Pi_{1} = - (p_{\theta_{1}} + \theta_{1} p),\qquad \Pi_{2} = -
(p_{\theta_{2}} + \theta_{2} p)
\end{equation}
which would satisfy (see Eq. (\ref{1}))
\begin{equation}
\Pi_{1}\star \Pi_{1} = - 2\kappa p = \Pi_{2}\star \Pi_{2},\qquad
\Pi_{1}\star \Pi_{2} = - \Pi_{2}\star \Pi_{1}
\end{equation}
Furthermore, through the graded Moyal bracket, they will lead to
covariant derivatives acting on any superfield on this space.

With these, let us consider a bosonic superfield $\Psi$ on this graded
manifold depending on $x,\theta_{1},\theta_{2}$. This $N=2$ superfield can,
of course, be decomposed and written as a sum of two $N=1$
superfields, but let us continue our discussion with $\Psi$. It is
known that there are only three nontrivial $N=2$ supersymmetrizations of
the KdV hierarchy which are integrable (corresponding to a parameter
$a=1,4, -2$). Let us consider the three cases separately. 

The Lax operator
\begin{equation}
L = p + \Pi_{1}^{-1}\star \Pi_{2}\star \Psi
\end{equation}
leads through the super Moyal-Lax equation
\begin{equation}
{\partial L\over \partial t} = \left\{L, \left(L^{3}\right)_{\geq
1}\right\}_{\kappa}
\end{equation}
to the equation
\begin{equation}
{\partial \Psi\over \partial t} = \left(-4\kappa^{2} \Psi_{xx} -
6\kappa (D_{1}D_{2}\Psi) \Psi + \Psi^{3}\right)_{x}
\end{equation}
which we recognize to be the $N=2$ supersymmetrization of the KdV
equation corresponding to $a=1$ \cite{6}.

On the other hand, the Lax operator
\begin{equation}
L = - \left(\Pi_{1}\star \Pi_{2} + \Psi\right)^{2}\label{20}
\end{equation}
leads, through the standard super Moyal-Lax equation,
\begin{equation}
{\partial L\over \partial t} = -{1\over 2\kappa} \left\{L,
\left(L^{3\over 2}\right)_{+}\right\}_{\kappa}
\end{equation}
to the equation
\begin{equation}
{\partial \Psi\over \partial t} = \left(-4\kappa^{4} \Psi_{xx} +
3\kappa^{2} (D_{1}\Psi) (D_{2}\Psi) + 6\kappa^{2} (D_{1}D_{2}\Psi)
\Psi + \Psi^{3}\right)_{x}
\end{equation}
which is the $N=2$ supersymmetrization of the KdV equation
corresponding to $a=4$ \cite{3}.

Finally, we note that if we take the Lax operator to be the $()_{\geq
1}$ projection of that in Eq. (\ref{20}), namely,
\begin{equation}
{\cal L} = \left(L\right)_{\geq 1}
\end{equation}
then, the standard super Moyal-Lax equation
\begin{equation}
{\partial {\cal L}\over \partial t} = {1\over \kappa} \left\{{\cal
L},\left({\cal L}^{3\over 2}\right)_{+}\right\}_{\kappa}
\end{equation}
leads to the equation
\begin{equation}
{\partial \Psi\over \partial t} = \left(8\kappa^{4} \Psi_{xx} -
6\kappa^{2} (D_{1}\Psi) (D_{2}\Psi) + \Psi^{3}\right)_{x}\label{b}
\end{equation}
which is the $N=2$ supersymmetrization of the KdV equation
corresponding to the parameter $a=-2$ \cite{3}.

Thus, we see that all three of the $N=2$ supersymmetrizations of the
KdV equation can be given a super Moyal-Lax
representation. We can carry over the arguments of section {\bf 3} and
show that all the three super Moyal-Lax equations can be given the
meaning of Hamiltonian equations and can be derived from suitable
actions. Furthermore, the dispersionless limits of these models
can be obtained by taking the $\kappa\rightarrow 0$
limit. Surprisingly, all three models, in this limit, yield
\begin{equation}
{\partial \Psi\over \partial t} = \left(\Psi^{3}\right)_{x}\label{I}
\end{equation}
whose bosonic limit is
\begin{eqnarray}
{\partial\omega\over \partial t} & = &
\left(\omega^{3}\right)_{x}\nonumber\\
{\partial u\over \partial t} & = & 3 \left(u\omega\right)_{x}
\end{eqnarray}
where we have identified $\Psi = \omega + \theta_{1}\psi_{1} +
\theta_{2}\psi_{2} + 
\theta_{2}\theta_{1} u$ and this equation contains the dispersionless
mKdV equation (in the limit $u=0$). Thus, Eq. (\ref{I}), can be
thought of as a trivial supersymmetrization of the dispersionless mKdV
equation (in the sense that it does not contain supersymmetric
covariant derivatives).

We can, in fact, give a Lax description for this supersymmetrization
of the mKdV equation \cite{20} in the following way. First, let us
consider  a
rotated  basis and define two covariant derivatives as
\begin{equation}
D_{1} = \partial_{\theta_{1}} - {1\over 2} \theta_{2}
\partial_{x},\qquad D_{2} = \partial_{\theta_{2}} - {1\over 2}
\theta_{1} \partial_{x}
\end{equation}
Unlike the conventional ones, these covariant derivatives satisfy
\begin{equation}
D_{1}^{2} = 0 = D_{2}^{2},\qquad D_{1}D_{2}+D_{2}D_{1} = -
\partial_{x}
\end{equation}
In this case, we can define
\begin{equation}
\Pi_{1} = - p_{\theta_{1}} + {1\over 2} \theta_{2} p,\qquad \Pi_{2} =
- p_{\theta_{2}} + {1\over 2} \theta_{1} p
\end{equation}
such that, with the star product defined in Eq. (\ref{1}), we have
\begin{equation}
\Pi_{1}\star \Pi_{1} = 0 = \Pi_{2}\star \Pi_{2},\qquad
\left\{\Pi_{1},\Pi_{2}\right\}_{\kappa} = p
\end{equation}
Furthermore, it can be checked that $\Pi_{1,2}$, through the super
Moyal bracket, generate appropriate covariant derivatives in the
rotated basis.

With these operators, let us next define
\begin{equation}
L = p^{2} + \Psi\star p + (D_{2}\Psi)\star \Pi_{1}\label{J}
\end{equation}
Then, it is easy to check that the super Moyal-Lax equation
\begin{equation}
{\partial L\over \partial t} = -8 \left\{L, \left(L^{3\over
2}\right)_{\geq 1}\right\}_{\kappa}
\end{equation}
leads to
\begin{equation}
{\partial \Psi\over \partial t} = \left(-8\kappa^{2} \Psi_{xx} +
12\kappa (D_{1}\Psi)(D_{2}\Psi) + \Psi^{3}\right)_{x}
\end{equation}
This equation can be compared with Eq. (\ref{b}) (recall, however,
that the covariant derivatives in the two equations correspond to
different basis). In fact, we note
here that the dispersionless limit of this equation, 
\begin{equation}
{\partial \Psi\over \partial t} = \left(\Psi^{3}\right)_{x}\label{c}
\end{equation}
can be described by a Lax equation of the following form. Let us
consider
\begin{equation}
L = p^{2} + \Psi p\label{K}
\end{equation}
Then, the ordinary Poisson bracket equation,
\begin{equation}
{\partial L\over \partial t} = -8 \left\{L, \left(L^{3\over
2}\right)_{\geq 1}\right\}
\end{equation}
leads to Eq. (\ref{c}). It is not obvious that the Lax operator
(\ref{J}) reduces exactly to that in Eq. (\ref{K}). However, that this
is true can be seen in the following manner. The linear equation
associated with the Lax operator in Eq. (\ref{J}) has the form
\begin{equation}
L\star \psi = \lambda \psi\label{L}
\end{equation}
where $\lambda$ represents the spectral parameter. Both the
eigenfunction and the spectral parameter are functions of $\kappa$,
the deformation parameter. Therefore, they can be expanded in a power
series in $\kappa$, as, say
\begin{equation}
\psi = \psi_{0} + \kappa \psi_{1} + \kappa^{2} \psi_{2} +\cdots
\end{equation}
It can be determined easily that $\psi_{0}$, the component which survives in
the dispersionless limit ($\kappa\rightarrow 0$), has the form
$\psi_{0}=\Pi_{1} \phi$. As a result of this in the dispersionless
limit, the last term of the Lax operator (\ref{J}) drops out in the
linear equation (\ref{L}) (recall that in the dispersionless limit,
$\kappa\rightarrow 0$, we have $\Pi_{1}^{2}=0$) so that the Lax operator
in (\ref{K}) truly represents the reduction of Eq. (\ref{J}) in the
dispersionless limit.

Let us also note here that Eq. (\ref{c}) can be checked to have at
least three Hamiltonian structures of the forms
\begin{equation}
{\cal D}_{1} = \partial,\quad {\cal D}_{2} = \partial \Psi +
\Psi\partial,\quad {\cal D}_{3} =
\partial\Psi\partial^{-1}\Psi\partial
\end{equation}
each of which can be checked to satisfy Jacobi identity.

\section{Conclusion:}

We have generalized our earlier discussion of Moyal-Lax representation
for bosonic integrable systems \cite{13} to supersymmetric ones, with
simple  as
well as extended supersymmetries. We have derived various properties
of the supersymmetric star product. Within the context of the $N=1$
supersymmetric KdV equation, we have shown how the parameter of
deformation, in such systems, is related to the central charge of the
second Hamiltonian structure. We have also shown how the super
Moyal-Lax equation can be interpreted as a Hamiltonian equation and
can be derived from an action, much like in the bosonic case. The
conserved charges as well as the first two Hamiltonian structures are
constructed. We show how one can take the dispersionless limit of this
model within the super Moyal-Lax representation, the limit being
singular. (It is for this reason that the standard construction of a
Lax description for such dispersionless systems fails and one needs an
alternate description \cite{16}-\cite{17}.) The conserved quantities
as well as the Hamiltonian structures also reduce to the corresponding
quantities of the dispersionless models. This clarifies why the
construction of the first Hamiltonian structure for the dispersionless
sKdV system had failed so far. We have also briefly discussed the
super Moyal-Lax representations for the $N=1$ supersymmetric two
boson equation, non-linear Schr\"{o}dinger equation as well as the
modified KdV equation. We have also discussed the super Moyal-Lax
representations for the various $N=2$ supersymmetrizations of the KdV
equation. The dispersionless limits of these systems and their
properties are also discussed. 
  
This work was supported in part by US DOE grant No. DE-FG 02-91ER40685 as
well as by NSF-INT-

\end{document}